\documentclass{article}
\input epsf

\usepackage{color}

\begin{document}
\centerline{\bf A Spectroscopic Study of Kepler Asteroseismic Targets
\footnote{Based in part on data obtained with the ESA Hipparcos satellite. 
The ground-based data used in this paper have been obtained at the {\it 
M.G. Fracastoro\/} station of the Catania Astrophysical Observatory, the 
Oak Ridge Observatory, Harvard, Massachusetts, and the F.L.\ Whipple 
Observatory, Mount Hopkins, Arizona.}}

\vspace{0.3cm}
\centerline{by}

\vspace{0.3cm}
\centerline{J.~M o l e n d a - \.Z a k o w i c z$^1$, A.~F r a s c a$^2$, 
D.W.~L a t h a m$^3$ and M.~J e r z y k i e w i c z$^1$}

\vspace{0.3cm}
\centerline{$^1$ Astronomical Institute, University of  Wroc{\l }aw, Kopernika 11,}
\centerline{51-622 Wroc{\l }aw, Poland, e-mail: (molenda,mjerz)@astro.uni.wroc.pl}
\vspace{0.1cm}
\centerline{$^2$ Catania Astrophysical Observatory, Via S.Sofia 78,} 
\centerline{95123 Catania, Italy, e-mail: afr@oact.inaf.it}
\vspace{0.1cm}
\centerline{$^3$ Harvard-Smithsonian Center for Astrophysics, 60 Garden Street,} 
\centerline{Cambridge, MA 02138, USA, e-mail: dlatham@cfa.harvard.edu}

\vspace{0.5cm}
\centerline{\it Received ...}

\vspace{1cm}
\centerline{ABSTRACT}

\vspace{0.5cm} 
{\small
Reported are spectroscopic observations of 15 candidates for Kepler primary 
asteroseismic targets and 14 other stars in the Kepler field, carried out at three 
observatories (see the footnote). For all these stars, the radial 
velocities, effective temperature, surface gravity, metalicity, 
and the projected rotational velocity are derived from two separate 
sets of data by means of two independent methods. In addition, MK type 
is estimated from one of these sets of data. 

Three stars, HIP\,94335, HIP\,94734, and HIP\,94743, are found to have 
variable radial-velocity. For HIP\,94335 = FL Lyr, a well-known 
Algol-type eclipsing variable and a double-lined spectroscopic binary, 
the orbital elements computed from our data agree closely with those of 
Popper et al. For HIP\,94734 and HIP\,94743 = V\,2077 Cyg, which we 
discover to be single-lined systems, orbital elements are derived. In 
addition, from our value of the orbital period and the Hipparcos epoch 
photometry, HIP\,94743 is demonstrated to be a detached eclipsing binary. 

}

\vspace{0.5cm}
\centerline{\bf 1. Introduction}

\vspace{0.5cm} 
Kepler\footnote{http://kepler.nasa.gov/} is a NASA space mission, 
scheduled for launch in February 2009. The Kepler equipment consists of 
a Schmidt telescope, with a 1.4-m primary mirror and a 0.95-m corrector 
plate, and a 430--890 nm FWHM bandpass  photometer, featuring an array 
of 42 CCDs, 50$\times$25 mm and 2200$\times$1024 pixels each. The dynamic 
range is 9--15 mag. Kepler will observe pre-selected stars in a 12-degree 
field in the Cygnus-Lyra region. The same field will be observed for the 
entire life-time of the Kepler mission, i.e., at least 3.5 years. The main 
scientific goal is the detection of Earth-size and larger planets by means 
of the method of photometric transits (Borucki et al.~1997). 

Kepler photometry will also be used for detecting solar-like pulsations 
in program stars and deriving accurate values of their pulsation 
frequencies, allowing an 
investigation of internal structure of these stars by means of asteroseismology 
(see, e.g., Christensen-Dalsgaard 2004). This investigation will be
carried out by members of KASC\footnote{http://astro.phys.au.dk/KASC/ 
Kepler Asteroseismic Science Consortium (KASC) is an international 
consortium of researchers dedicated to the asteroseismic analysis of 
Kepler data. It is a part of the Kepler Asteroseismic Investigation, 
which is coordinated by the Department of Physics and Astronomy, 
University of Aarhus, Denmark.} under the lead of J.\ 
Chri\-sten\-sen-Dalsgaard from the Department of Physics and Astronomy of 
the Aarhus University.

Molenda-\.Zakowicz et al.~(2006) have provided a list of 104 Hipparcos stars that were 
selected as candidates for asteroseismic targets for Kepler. These authors list 29 
candidates for primary asteroseismic targets, one being a binary whose components are 
treated separately by Molenda-\.Zakowicz et al.~(2006) but as an individual object in this 
paper. Since 14 of these stars fall either just beyond the Kepler CCD chips or into star 
tracker corners, and are not expected to be observed in normal conditions, in the present 
paper we narrow down the definition of the primary asteroseimic targets (hereafter, PATS) 
to 15 stars that fall onto active chips of Kepler CCDs.

We expect all these stars to show solar-like oscillations with 
amplitudes possible to be detected in Kepler photometry. Having the 
oscillation frequencies measured, the internal structure of these 
objects will be examined by means of the asteroseismic analysis. The 
results of this study, particularly the values of the stellar radii, 
will be used by the Kepler team for determining the radii of planets
and the parameters of the planetary systems, providing that such will 
be discovered.

Since the model computations require the effective temperature, surface gravity, 
metalicity, and the projected rotational velocity to be input parameters, in 2005 we 
started a program of ground-based observations, aiming at the determination of these 
parameters. In this paper, we present the results of our spectroscopic study, obtained for 
all the 29 stars. 

After giving an account of the spectroscopic observations and reductions in Sect.\ 2, in 
Sect.\ 3 we discuss three stars showing variable radial velocity, one known Algol-type 
eclipsing variable and a double-lined spectroscopic binary, HIP\,94335 = FL Lyr, and two 
single-lined spectroscopic binaries, HIP\,94734 and HIP\,94743 = V\,2077 Cyg, discovered 
in the present work. For HIP\,94335 we find the orbital elements computed in this paper to 
be in a very good agreement with those of Popper et al.\ (1986). For HIP\,94734 and 
HIP\,94743, we derive orbital elements for the first time. In addition, using our value of 
the orbital period and the Hipparcos epoch photometry (ESA 1977), we demonstrate that 
HIP\,94743 is a detached eclipsing binary. In Sect.\ 4, we determine the effective 
temperature, surface gravity, metalicity, and MK spectral type of the 29 stars; in 
Sect.\ 5, we give their projected rotational velocity. Sect.\ 6 contains a summary.

\vspace{0.5cm}
\centerline{\bf 2. Observations and Reductions}

\vspace{0.5cm}
The observations were carried out at the {\it M.G. Fracastoro\/} station 
(Serra La Nave, Mount Etna, elevation 1750 m) of the Catania Astrophysical 
Observatory (CAO), Italy (74 spectrograms), at the Oak Ridge Observatory 
(ORO), Harvard, Massachusetts (32 spectrograms), and at the F.L.\ Whipple 
Observatory (FLWO), Mount Hopkins, Arizona (126 spectrograms). 

At CAO, we used a 91-cm telescope and the fiber-fed echelle spectrograph 
FRESCO. The spectra were recorded with resolving power R=21\,000 in a wide 
spectral range that covered about 2\,500 {\AA} in 19 orders. As the 
detector, we used a thinned back-illuminated CCD SITe chip (SI033B) with 
1024x1024 24x24-$\mu$m pixels. 

The telescope used at ORO was the 1.5-m Wyeth reflector. At FLWO, two 
telescopes were used, viz., the 1.5-m Tillinghast reflector and the 
Multiple Mirror Telescope (before it was converted to the monolithic 6.5-m 
mirror). The spectra were obtained by means of echelle spectrographs 
with resolving power R=35\,000 (CfA Digital Speedometers). The detector 
consisted of an intensified  photon-counting Reticon. In this system, a 
single 45 {\AA} spectrogram, centered at $\lambda \simeq$ 5187 {\AA}, was 
recorded in one exposure. 

The spectrograms measured with FRESCO were reduced with the use of IRAF\footnote{IRAF is 
distributed by the National Optical Astronomy Observatory, which is operated by the 
Association of Universities for Research in Astronomy, Inc.}. The spectra were extracted 
with the {\sf apall} task and the radial velocities were determined {\em via\/} the 
cross-correlation method, also provided by IRAF in the {\sf fxcor} task.  Arcturus, 
$\beta$ Oph, 54 Aql, 31 Aql or $\gamma$ Aql, for which precise values of R.V.\ are 
available (Udry et al.~1999), served as the radial-velocity templates. These stars were 
observed on the same nights as the program stars. Care was taken to match spectral types 
of the template and program star as closely as possible. 

\begin{table}
\begin{center}
\centerline{T a b l e \quad 1}

{15 G-type dwarfs used for calculating the offset between radial 
velocities measured with FRESCO, R.V.(F), and those measured 
with the CfA Speedometers, $\rm R.V. (CfA)$. The last three columns contain 
the number and time-span of observations made with the CfA Speedometers, 
$\rm N_{CfA}$ and Span (CfA), and 
the difference between the R.V. values measured with FRESCO and the
CfA Speedometers, $\rm \Delta R.V.  = R.V.(F) - R.V.(CfA)$}

\vspace{0.3cm}
{\small
\begin{tabular} {rrrrrrrr} \\
\hline\noalign{\smallskip}
HD    & R.V.(F)& s.e. & R.V.(CfA) & s.e. &$\rm N_{CfA}$ & 
Span (CfA) & $\rm \Delta R.V.$ \\
\noalign{\smallskip}\hline\noalign{\smallskip}

 90839 & +10.13 & 0.25 &  +8.39 & 0.46 &  67 & 9900 & +1.74\\
 99984 & -32.26 & 0.33 & -33.00 & 0.48 &  18 & 9883 & +0.74\\
102870 &  +5.13 & 0.24 &  +4.25 & 0.46 & 298 &10072 & +0.88\\
107213 &  -9.56 & 0.37 &  -9.61 & 0.56 &  32 & 7945 & +0.05\\
114710 &  +6.58 & 0.22 &  +5.22 & 0.52 & 150 & 9048 & +1.36\\
123782 & -15.14 & 0.25 & -14.12 & 0.76 &  33 & 5967 & -1.02\\
136064 & -48.12 & 0.30 & -48.44 & 0.44 &  17 & 9207 & +0.32\\
142373 & -56.10 & 0.27 & -56.39 & 0.44 &  41 &10303 & +0.29\\
161096 & -11.91 & 0.12 & -12.32 & 0.44 & 113 & 5596 & +0.41\\
185144 &  26.04 & 0.22 & +26.05 & 0.47 &   6 & 2983 & -0.01\\
186408 & -28.07 & 0.25 & -27.38 & 0.44 &  16 & 5141 & -0.69\\
186427 & -27.79 & 0.16 & -28.13 & 0.53 &  23 & 7122 & +0.34\\
187691 &  +0.55 & 0.24 &  -0.15 & 0.40 & 517 & 9269 & +0.70\\
201891 & -44.68 & 0.26 & -44.78 & 0.56 &  25 & 6694 & +0.10\\
219623 & -25.97 & 0.25 & -27.29 & 0.49 &  15 & 8897 & +1.32\\
\noalign{\smallskip}\hline
\end{tabular}
}
\end{center}
\end{table}

Spectra measured at ORO and FLWO were extracted and rectified by means 
of a special procedure developed for these observatories and described
in detail in Latham et al.~(1992). For the templates, the grid of 
synthetic spectra, based on model atmospheres of R.L.\ Kurucz and 
computed by Jon Morse for this specific wavelength window and resolution 
was used (see Torres et al.~2002). 

Before the analysis of radial velocities was started, all measurements were moved to one 
reference system. We calculated the offset in radial velocity using the mean values of 
R.V.\ measured for 15 G-type dwarfs with FRESCO (single measurements, mean values over 19 
orders) and the CfA Digital Speedometers (long time-span of observations and tens or 
hundreds  measurements).  In Table 1, we list these 15 stars, their mean radial velocity 
in km\,s$^{-1}$ measured with FRESCO and CfA Speedometers, the standard deviation of the 
measurements, the number of CfA Speedometers' spectrograms, $\rm N_{CfA}$, the time-span 
of CfA observations in days, and the difference between the mean radial velocity values. 
We conclude that the FRESCO zero-point is stable in time, which allows us to combine the 
data sets, and that the unweighted average difference between the FRESCO and CfA 
zero-point is equal to +0.44$\pm$0.19. With this in mind, we merged the FRESCO and CfA 
Digital Speedometers' observations, subtracting 0.5 km\,s$^{-1}$ from each FRESCO's 
measurement.  All results that we discuss in this paper are obtained from the merged data.

\begin{table}
\begin{center}
\centerline{T a b l e \quad 2}

{Radial velocities (in km/s) of the program stars.\\
 The code in the last column indicate stars that fall\\
 onto the active chips of Kepler CCDs, ``A'', stars that\\
 fall into CCD gaps, ``G'', or stars that fall into star\\
 \hspace{6mm}tracker corners, ``S''.\\
The individual velocities are not corrected for the shift\\
\hspace{5mm}between observatories}

\vspace{0.1cm}
{\small
\begin{tabular}    
{lcrclc} \\
\hline\noalign{\smallskip}
HIP    & HJD-2400000  &R.V. & s.e. & Instrument & code\\
\noalign{\smallskip}\hline\noalign{\smallskip}
91128  & 45839.7031 &  $-$31.43 & 0.54 & Wyeth       &S\\
91128  & 45839.9676 &  $-$32.04 & 0.39 & Tillinghast &S\\
91128  & 53512.8557 &  $-$32.15 & 0.37 & Tillinghast &S\\
91128  & 53517.8486 &  $-$32.05 & 0.36 & Tillinghast &S\\
91128  & 53894.3541 &  $-$31.47 & 0.56 & FRESCO      &S\\
   &     &         &           &               \\
92922  & 51784.6670 &  $-$39.68 & 0.71 & Wyeth       &G\\ 
92922  & 53512.8595 &  $-$40.03 & 0.26 & Tillinghast &G\\ 
92922  & 53517.8522 &  $-$39.82 & 0.35 & Tillinghast &G\\ 
92922  & 53549.3790 &  $-$39.00 & 0.23 & FRESCO      &G\\ 
92922  & 53555.4006 &  $-$38.58 & 0.19 & FRESCO      &G\\ 
   &     &         &          &             \\
93011  & 53512.8629 &  $-$23.43 & 0.90 & Tillinghast &A\\
93011  & 53517.8551 &  $-$23.69 & 0.88 & Tillinghast &A\\
93011  & 53552.3660 &  $-$21.74 & 0.92 & FRESCO      &A\\
93011  & 53558.5257 &  $-$21.62 & 0.71 & FRESCO      &A\\
   &     &         &          &             \\
94145  & 53512.8669 &    0.76 & 0.40 & Tillinghast   &G\\
94145  & 53517.8586 &    3.41 & 0.49 & Tillinghast   &G\\
94145  & 53543.9805 &    1.32 & 0.53 & Tillinghast   &G\\
94145  & 53549.4241 &    2.25 & 0.24 & FRESCO        &G\\
94145  & 53553.5995 &    3.04 & 0.29 & FRESCO        &G\\
94145  & 53659.6048 &    3.56 & 0.63 & Tillinghast   &G\\
94145  & 53895.3536 &    2.23 & 0.72 & FRESCO        &G\\
\multicolumn{6}{c}{\dotfill} \\         
99267  & 53905.4480 & $-$193.98 & 1.65 & FRESCO      &S\\
\noalign{\smallskip}\hline    
\end{tabular}
}
\end{center}
\end{table}

In Table 2, we give the individual radial-velocity measurements not 
corrected for this offset. The table is available in electronic form 
from the Acta Astronomica Archive (see the cover page). A sample, 
containing the heading, the first 21 rows, and the last row, is printed 
below. In the first column we give the HIP number, in the second, the 
Heliocentric Julian Day of the middle of the exposure, in the third and 
fourth, the radial velocity, R.V., and the standard error, s.e., then, 
the instrument used, and in the last column, the information 
whether the star falls onto the active pixels of Kepler CCDs, coded with 
``A'', into the gaps between CCD chips, coded with ``G'', or into a star 
tracker corner, coded with ``S''. 

\begin{table}
\begin{center} \centerline{T a b l e \quad 3}
{15 PATS and the remaining 14 stars: mean radial velocity, R.V., in km\,s$^{-1}$,\\ 
the ratio of 
external-to-internal error, e/i, the sum of the residuals divided by the 
internal error estimate squared, $\chi ^2$, and the probability that a 
star with constant velocity will have $\chi ^2$ values larger than the 
observed one, $\rm P(\chi ^2)$, or the classification to the type of 
spectroscopic variability
}

\vspace{0.3cm}
{\small
\begin{tabular} {cccrrrcrrr} \\
\hline\noalign{\smallskip}
HIP   & $\alpha _{\rm 2000}$ & $\delta _{\rm 2000}$ & 
$N$ & span  & R.V. & s.e. &
e/i & $\chi ^2$ & $\rm P(\chi ^2)$\\
\noalign{\smallskip}\hline\noalign{\smallskip}
\multicolumn{10}{c}{PATS}\\
\noalign{\smallskip}\hline\noalign{\smallskip}
 93011 & 18 56 53.36 & +48 41 38.4 &  4 &   46 &  -22.87 & 0.44 & 0.9 &   2.6 & 0.46 \\
 94335 & 19 12 04.86 & +46 19 26.5 & 14 &  181 &  -38.32 & 0.25 & 58.7& 3293.6& SB2 \\
 94497 & 19 13 58.83 & +39 50 37.4 &  5 & 1772 &   +7.64 & 0.19 & 0.8 &   2.6 & 0.63 \\
 94565 & 19 14 45.00 & +51 08 42.4 &  6 &  916 &  -10.82 & 0.20 & 1.1 &   2.8 & 0.42 \\
 94734 & 19 16 34.89 & +40 02 48.8 & 16 &  916 &  -25.46 & 1.01 & 7.0 & 50.2& SB1 \\
 95098 & 19 20 47.91 & +44 09 18.7 &  5 &  916 &  +10.41 & 0.19 & 0.4 &   0.4 & 0.94 \\
 95637 & 19 27 13.74 & +50 50 56.5 &  4 &   46 &  -11.61 & 0.40 & 0.7 &   1.7 & 0.65 \\
 95733 & 19 28 21.01 & +39 04 50.6 &  5 &  917 &  +20.52 & 0.35 & 0.6 &   1.6 & 0.79 \\
 96634 & 19 38 51.16 & +43 52 28.2 &  5 & 1769 &  +15.25 & 0.27 & 0.4 &   1.2 & 0.88 \\
 96735 & 19 39 52.12 & +45 09 33.0 &  5 & 1299 &  -16.64 & 0.19 & 0.6 &   1.4 & 0.85 \\
 97219 & 19 45 30.09 & +50 46 18.8 & 12 & 4331 &  -20.65 & 0.16 & 1.1 &  12.8 & 0.31 \\
 97337 & 19 47 10.19 & +40 20 25.0 &  1 &    0 &  -70.50 & 0.49 & 0.0 &   0.0 & 1.00 \\
 97657 & 19 50 50.14 & +48 04 49.1 &  7 & 1996 &  -63.55 & 0.15 & 0.8 &   4.6 & 0.60 \\
 97974 & 19 54 41.18 & +42 25 50.5 &  2 &   47 &  -31.75 & 0.34 & 0.9 &   0.7 & 0.39 \\
 98655 & 20 02 17.06 & +45 46 58.0 &  4 &  396 &  -52.20 & 0.28 & 1.2 &   4.6 & 0.20 \\
\noalign{\smallskip}\hline\noalign{\smallskip}
\multicolumn{10}{c}{The remaining stars}\\
\noalign{\smallskip}\hline\noalign{\smallskip}
 91128 & 18 35 18.01 & +45 44 35.4 &  5 & 8055 &  -31.93 & 0.23 & 0.6 &   1.1 & 0.90 \\
 92922 & 18 55 54.82 & +41 31 17.7 &  5 & 1771 &  -39.62 & 0.20 & 0.8 &   4.6 & 0.33 \\
 94145 & 19 09 49.51 & +44 33 26.9 &  8 &  386 &   +2.36 & 0.40 & 1.8 &  26.8 & 0.00 \\
 94704 & 19 16 13.40 & +37 04 18.8 &  4 & 2031 & -347.30 & 0.39 & 0.6 &   0.9 & 0.83 \\
 94743 & 19 16 44.50 & +50 38 47.8 & 30 &  917 &   -0.40 & 0.66 & 24.7& 750.8 & SB1 \\
 94898 & 19 18 41.45 & +43 14 12.5 &  6 & 2122 &  -37.34 & 0.38 & 2.1 &  22.6 & 0.00 \\
 95631 & 19 27 05.78 & +46 27 14.5 &  6 & 1771 &  -24.34 & 0.20 & 1.0 &   9.7 & 0.08 \\
 95638 & 19 27 14.11 & +43 15 38.5 &  3 &  391 &  -32.04 & 0.54 & 2.0 &   7.4 & 0.03 \\
 95843 & 19 29 35.34 & +45 30 46.3 &  4 &   40 &  -22.02 & 0.27 & 0.8 &   1.6 & 0.67 \\
 96146 & 19 32 56.38 & +50 55 52.1 &  4 &   43 &  -29.45 & 0.72 & 1.8 &  10.0 & 0.02 \\
 97168 & 19 44 59.23 & +51 35 41.1 &  5 & 1679 &  -91.65 & 0.29 & 1.3 &   6.5 & 0.16 \\
 98381 & 19 59 19.33 & +42 10 06.4 &  5 & 1772 &  -73.14 & 0.24 & 1.1 &   4.2 & 0.38 \\
 98829 & 20 04 10.58 & +42 30 43.6 &  6 & 2574 &  -42.01 & 1.13 & 5.9 & 133.1 & 0.00 \\
 99267 & 20 09 01.32 & +42 51 52.0 & 46 & 8425 & -196.10 & 0.14 & 1.3 &  76.6 & 0.00 \\

\noalign{\smallskip}\hline
\end{tabular}
}
\end{center}
\end{table}

In Table 3, we list the 15 PATS and the 14 remaining stars, 
their mean radial velocity,
R.V., in km\,s$^{-1}$, the ratio of external-to-internal error, e/i, 
the sum of the residuals divided by the internal error estimate squared, 
$\chi ^2$, and the probability that a star with constant velocity will 
have $\chi ^2$ values larger than the observed one, $\rm P(\chi ^2)$, or,
for the double stars, the classification to the type of spectroscopic 
binarity. For detailed description of the method of computing the respective 
errors, we refer to Latham et al.~(2002).

\vspace{0.5cm}
{\bf 3. Program Stars with Variable Radial-$\!$Velocity}

\vspace{0.5cm}
{\em 3.1. HIP\,94335 = FL Lyr}

\vspace{0.5cm}
In Table 2, there are 14 R.V.\ measurements of FL Lyr per component, twelve Tillinghast 
and two FRESCO. In the table, the primary component is denoted HIP\,94335-1, and the 
secondary component is denoted HIP\,93335-2. An orbital solution, obtained from the merged 
data (corrected for the offset between our observatories), with equal weights assigned to 
all measurements, is listed in Table 4. A phase diagram, computed using this solution, 
is shown in Fig.\ 1. 

\begin{table}
\begin{center}
\centerline{T a b l e \quad 4}
\vspace*{0.1cm}
{Orbital elements of HIP\,94335 = FL Lyr}

\begin{tabular}    
{l} \\
$P_{\rm orb} = 2.178097 \pm 0.000058$ d \\
$\gamma = -38.36 \pm 0.27$ km\,s$^{-1}$ \\
$K_1 = 93.54 \pm 0.34$ km\,s$^{-1}$ \\
$K_2 = 117.71 \pm 1.21$ km\,s$^{-1}$ \\
$e = 0.0063  \pm $ 0.0037 \\
$\omega_1 = 112^o \pm 47^o$\\
$T = {\rm HJD}\,2453595.10 \pm 0.28$ \\
$a_1 \sin i = (2.802 \pm 0.010) \times 10^6$ km\\
$a_2 \sin i = (3.525 \pm 0.036) \times 10^6$ km\\
{\sf M}$_1 \sin^3 i = 1.190 \pm 0.020$ M$_\odot$ \\
{\sf M}$_2 \sin^3 i = 0.944 \pm 0.012$ M$_\odot$ \\
\end{tabular}
\vspace{0.3cm}
\end{center}
\end{table}
\begin{figure} 
\epsfbox{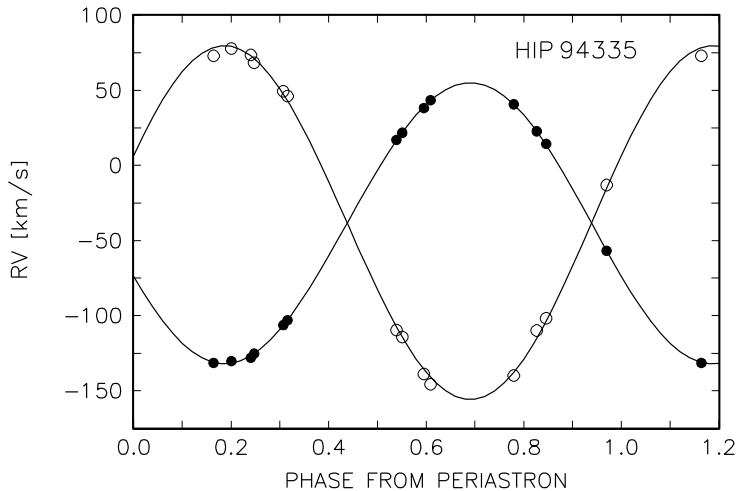}
\caption{The radial-velocity observations of HIP\,94335 = FL Lyr from Table 2 corrected 
for the shift between observatories. The radial-velocities of the primary component are 
shown with filled circles, and those of the secondary component, with open circles. The 
lines represent the orbital solution given in Table 4.}  
\end{figure}

A comparison of the numbers from Table 4 with the orbital elements of 
FL Lyr given in Table XVI of Popper et al.\ (1986) shows that (1) the 
two solutions are of similar accuracy, and (2) in no case the difference 
between the corresponding elements exceeds one standard deviation. 

There is also a very good agreement between the value of the orbital period given by 
Popper et al.\ (1986) in their Table VII and the value obtained by dividing the difference 
of the epoch of our R.V.\ changing from recession to approach and their epoch of the 
primary (photometric) minimum by the number of cycles that had elapsed between the two 
epochs. The difference between these two values of the orbital period amounts to $(6 \pm 
5) \times 10^{-7}$ d.

\vspace{0.5cm}
{\em 3.2. HIP\,94734}

\vspace{0.5cm}
For this star we have 16 measurements (see Table 2). The highest peak in the periodogram 
of these data occurs near 0.01 c/d. We conclude that HIP\,94734 is an SB1 system with 
orbital period close to 100 d. Assigning equal weights to all points, we derived the 
orbital elements given in Table 5. A phase diagram, computed using these elements is 
shown in Fig.\ 2. 

\begin{table}
\begin{center}
\centerline{T a b l e \quad 5}
\vspace*{0.1cm}
{Orbital elements of HIP\,94734}

\begin{tabular}    
{l} \\
\hspace{5mm}$P_{\rm orb} = 98.92 \pm 0.17$ d \\
\hspace{0.5cm}$\gamma = -25.47 \pm 0.45$ km\,s$^{-1}$ \\
\hspace{0.5cm}$K_1 = 16.18 \pm 0.75$ km\,s$^{-1}$ \\
\hspace{0.5cm}$e = 0.195 \pm 0.025$ \\
\hspace{0.5cm}$\omega_1 = 132.\!^o3 \pm 9.\!^o9$ \\
\hspace{0.5cm}$T = {\rm HJD}\,2454008.8 \pm 2.6$ \\
\hspace{0.5cm}$a_1 \sin i = (2.16 \pm 0.10) \times 10^7$ km\\
\hspace{0.5cm}$f(M) = 0.0410 \pm 0.0057$ M$_\odot$ \\
\end{tabular}
\vspace{0.3cm}
\end{center}
\end{table}

\begin{figure} \epsfbox{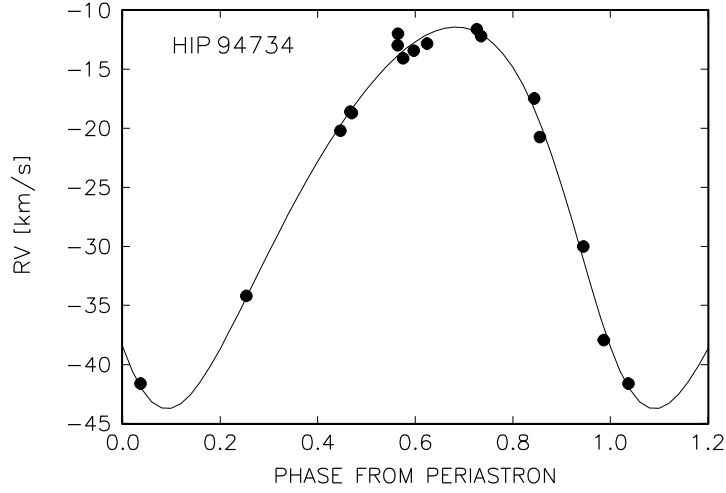}
\caption{The radial-velocity observations of HIP\,94734 from Table 2 
corrected for the shift between observatories 
(circles). The line represents the orbital solution given in Table 5.}  
\end{figure}

\vspace{0.5cm}
{\em 3.3. HIP\,94743 = V\,2077 Cyg}

\vspace{0.5cm}
This star was classified as a suspected eclipsing binary, E:, by  Kazarovets et 
al.~(1999) on the basis of Hipparcos photometry (ESA 1997). However, no period is 
reported in the Hipparcos catalogue. A periodogram of our 29 radial-velocity 
observations of the star (see Table 2) is dominated by a peak at 0.1684 d$^{-1}$. A 
sine-curve of this frequency fits the data satisfactorily. We conclude that HIP\,94743 
is an SB1 system with orbital period close to 5.938 d and small eccentricity. 
Assigning equal weights to all points, we derived the orbital elements given in Table 6. 
A phase diagram, computed using these elements, is shown in Fig.\ 3. 

\begin{figure} 
\epsfbox{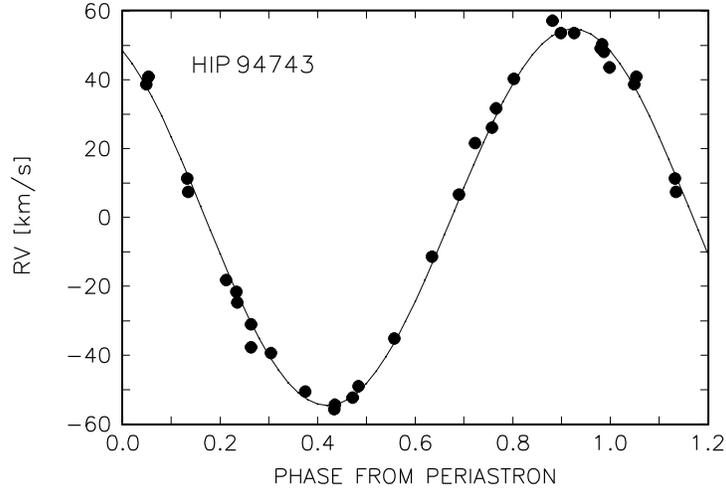} 
\caption{The radial-velocity curve of HIP\,94743 = V\,2077 Cyg from Table 2 
corrected for the shift between observatories 
(circles). The line represents the orbital solution given in Table 6.}  
\end{figure}

\begin{table}
\begin{center}
\centerline{T a b l e \quad 6}
\vspace*{0.1cm}
{Orbital elements of HIP\,94743 = V\,2077 Cyg}

\begin{tabular}    
{l} \\
$P_{\rm orb} = 5.93733 \pm 0.00025$ d \\
$\gamma = -0.42 \pm 0.54$ km\,s$^{-1}$ \\
$K_1 = 54.73 \pm 0.69$ km\,s$^{-1}$ \\
$e = 0.010 \pm 0.012$ \\
$\omega_1 = 28^o \pm 87^o$ \\
$T = {\rm HJD}\,2453952.7 \pm 1.4$ \\
$a_1 \sin i = (4.468 \pm 0.056) \times 10^6$ km\\
$f(M) = 0.1011 \pm 0.0038$ M$_\odot$ \\
\end{tabular}
\vspace{0.3cm}
\end{center}
\end{table}

Hipparcos magnitudes of the star phased with $P_{\rm orb} = 5.93733$ d show two minima of 
unequal depths, separated in phase by 0.5. The epoch of the primary minimum is shifted 
from the epoch of our R.V.\ changing from recession to approach by 0.06 in phase. Taking 
into account the fact that some 885 cycles have elapsed between the mid-epochs of the 
Hipparcos and our data, we get a slightly improved value of the orbital period, $P_{\rm 
orb} = 5.93720 \pm 0.00013$ d. The Hp magnitudes of HIP\,94743 are plotted as a function of 
phase of this period in Fig.\ 4. As can be seen from the figure, the system is widely 
detached. 

\begin{figure} 
\epsfbox{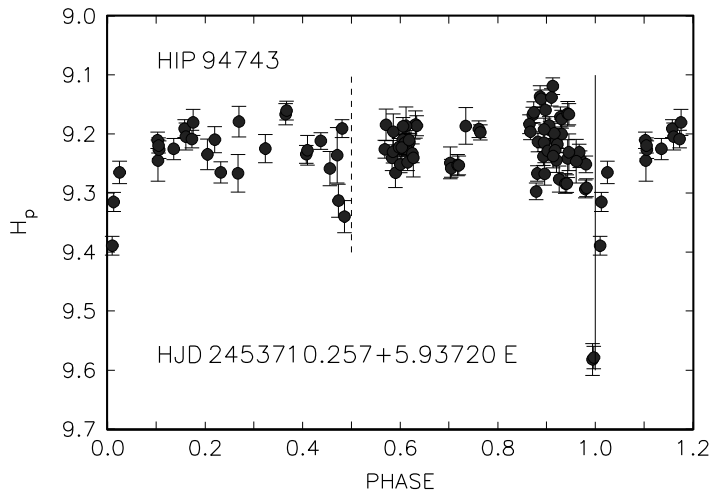} 
\caption{The Hp magnitudes of HIP\,94743 = V\,2077 Cyg (circles with error bars). The 
phases were computed with the ephemeris shown. Phases 0 and 0.5 
are indicated by the solid and dashed vertical line, respectively.}  
\end{figure}

\vspace{0.5cm}
{\em 3.4. Two Unconfirmed Spectroscopic Binaries}

\vspace{0.5cm} 
Of the remaining stars, HIP\,97219 = HD\,187055 was reported to be 
variable in radial velocity by Nordstr\"{o}m et al.~(2004), and 
HIP\,99267 = G125-64 was suspected to be a spectroscopic binary by 
Carney \& Latham (1987). Nordstr\"{o}m et al.~(2004) based their 
conclusion on four Coravel measurements. Our 12 radial-velocities of 
HIP\,97219 include eight Wyeth measurements. As can be seen from Table 2, 
the (external) standard deviation of the latter data and the mean value 
of their s.e.\ are both equal to 0.5 km\,s$^{-1}$. In addition, the 
remaining measurements (three Tillinghast and one FRESCO) fall within the 
range of 1.6 km\,s$^{-1}$ defined by the Wyeth data. Clearly, the 
star's R.V.\ does not 
vary by more than 1.6 km\,s$^{-1}$. The $\rm P(\chi ^2)$ computed from our
observations is equal to 0.3 and the ratio of external-to-internal error
is equal to 1.1. Thus, we do not confirm the result of 
Nordstr\"{o}m et al.~(2004). However, our data may not exclude the case 
of a very eccentric orbit for which we covered only the flat portion of 
the R.V.\ curve.

For HIP\,99267 = G125-64 there are 46 R.V.\ measurements in Table 2. 
These data include 19 discovery observations of Carney \& Latham (1987). 
The time distribution of the data is not uniform: after the first 42 
measurements, which cover an interval of eight years, there is a gap of 
almost 14 years; the last four measurements span an interval 
of one year. Disregarding these four measurements, one gets a range of 
5.3 km\,s$^{-1}$ and a standard deviation of 1.1 km\,s$^{-1}$. Comparing 
the 
latter value with the s.e.\ (which range from 0.9 to 1.7 km\,s$^{-1}$) we 
conclude that better data are needed to confirm the binary status of 
the star. We note also that the above-mentioned range of 5.3 km\,s$^{-1}$ 
is defined by measurements with the second and third largest s.e.\ and that
although the $\rm P(\chi ^2)$ computed from our data is equal to zero, 
the ratio of external-to-internal error is equal to 1.3 only.

\vspace{0.5cm} 
{\bf 4. Effective Temperature, Surface Gravity, Metalicity, and the MK Type}

\vspace{0.5cm}
{\em 4.1. From a Comparison with Standard Stars}

\vspace{0.5cm}
We have determined $T_{\rm eff}$, $\log g$  and $\rm [Fe/H]$ of the 
program stars from the FRESCO spectrograms using ROTFIT, an IDL code 
developed by A.F.\ and his coworkers (see, e.g., Frasca et al.\ 2003, 
2006). The method used by ROTFIT is similar to that of Katz et al.~(1998) 
and Soubiran et al.~(1998). It consists in comparing the spectra of 
program stars with a library of spectra of reference stars. The aim is to 
find 10 reference stars that best reproduce the target spectrum.  The 
weighted means of their astrophysical parameters, $T_{\rm eff}$, $\log 
g$, and $\rm [Fe/H]$, are adopted as estimates of the astrophysical 
parameters of the program star.  The spectrograms have to be cleaned of 
instrumental effects and cosmic rays, and brought to the same spectral 
resolution, the same wavelength scale and the same continuum level. The 
spectrogram of each program star is compared order by order with the 
spectrograms of the reference stars by the method of least squares. Since 
the reference and the program stars may have different rotational 
velocities, in the process of comparison the reference spectra are 
broadened in a wide range of $v\sin i$ to find the best match. The value 
of  $\chi ^2$  is used as a measure of the similarity of spectra. 

As discussed by Katz et al.~(1998), Soubiran et al.~(1998), and Frasca et al.~(2006), this 
method allows simultaneous and fast determination of $T_{\rm eff}$, $\log g$, and $\rm 
[Fe/H]$ even from spectrograms of low signal-to-noise ratio (S/N) or moderate resolution. This 
is a clear advantage over the classical methods, based on fitting specific lines or 
spectral regions with synthetic spectra, that normally require spectrograms of high 
resolution and high S/N. Moreover, classical methods require {\it a priori\/} information 
on $T_{\rm eff}$ and $\log g$ to determine $\rm [Fe/H]$ because this parameter depends on 
the two others.

In order to adopt this method to our purpose we have compiled a list of 
240 slowly rotating reference stars ($v\sin i<15$ km\,s$^{-1}$) for which 
spectrograms are available from the ELODIE archive (Prugniel \& Soubiran 
2001). This archive contains high-resolution ($R$ = 42\,000) and high S/N 
spectrograms obtained with the Haute-Provence Observatory's 193-cm 
telescope and the fiber-fed echelle spectrograph ELODIE. For the majority 
of our reference stars, we adopted $T_{\rm{eff}}$, $\log g$, and $\rm 
[Fe/H]$ from the TGMET library of Soubiran et al.~(1998). For stars that 
are not included in this library, we adopted recent values of $T_{\rm 
{eff}}$, $\log g$, and $\rm [Fe/H]$ from the literature. In Table 7, 
available electronically from the Acta Astronomica archive, we list the 
240 stars and their $T_{\rm {eff}}$, $\log g$, and $\rm [Fe/H]$. For stars 
that are not included in the TGMET library, the references are given at 
the bottom of the table. As can be seen from the table, the values of 
$T_{\rm{eff}}$, $\log g$, and $\rm [Fe/H]$ of our ELODIE reference 
stars are in the ranges [3700\,K, 10800\,K], [0.2, 4.9], and [0.50, 
$-$2.38], respectively, wide enough to cover the expected ranges of the 
parameters of program stars. The last column of Table 7 contains the MK 
spectral type, compiled from the literature, with the Hipparcos Input 
Catalogue used as a guide. 

Since the ELODIE reference stars and the program stars were observed with 
different instruments and different resolutions, in the analysis we had 
to degrade the ELODIE spectrograms to match the lower resolution of FRESCO.
To check whether this procedure affects the resulting atmospheric 
parameters, we performed parallel computations with ROTFIT using  
reference stars observed with FRESCO. In order to save the observing time, 
we have chosen a less dense grid of slowly rotating reference stars 
with known astrophysical parameters. The list contains only 82 stars with  
$T_{\rm{eff}}$, $\log g$, and $\rm [Fe/H]$ in the ranges [3300\,K, 7700\,K],
[0.70, 4.55], and [0.35, $-$2.65], respectively. Since the  grid is 
coarse close to its edges, we have determined the atmospheric parameters 
only for these stars that fall into the ranges [4750\,K, 6750\,K], 
[3.8, 4.6], and [$-$0.5, 0.5] for $\log T_{\rm eff}$, $\log g$, and $\rm [Fe/H]$,
respectively, where the grid is dense enough to make our determinations 
reliable. In Table 8, available electronically from the Acta Astronomica 
Archive, we list the FRESCO reference stars and their adopted atmospheric 
parameters. As before, for the majority of stars we used the values from 
the TGMET library of Soubiran et al.~(1998), and for the remaining stars 
we adopted recent determinations from the literature, referenced at the 
bottom of the table. 

For each program star, we have selected five most similar ELODIE 
reference stars from Table 7 (those giving the lowest $\chi ^2$ values) 
and computed means of their $T_{\rm{eff}}$, $\log g$, and $\rm [Fe/H]$. 
The resulting values we consider to be representative for the program 
stars. They are listed in Table 9. In the last but one column of the table, 
we give MK spectral type assigned using a mean of MK types of two most 
similar reference stars selected in each order. We find our 
classification fully consistent with the relations between $T_{\rm{eff}}$ 
and spectral type given by Johnson (1966) for dwarfs and giants. 
For metal-deficient program stars no comparison was 
possible because Johnson (1966) does not consider such stars. In the last 
column of Table 9, we list spectral types taken from the Simbad database.

In addition, for 16 program stars that fall within the limits of the FRESCO 
grid we derived means of $T_{\rm{eff}}$, $\log g$, and $\rm [Fe/H]$ using 
reference stars from Table 8. In this case, we used three most similar 
reference stars for each program star. The results are given in Table 10. 

\begin{figure}
\begin{center}
\epsfbox{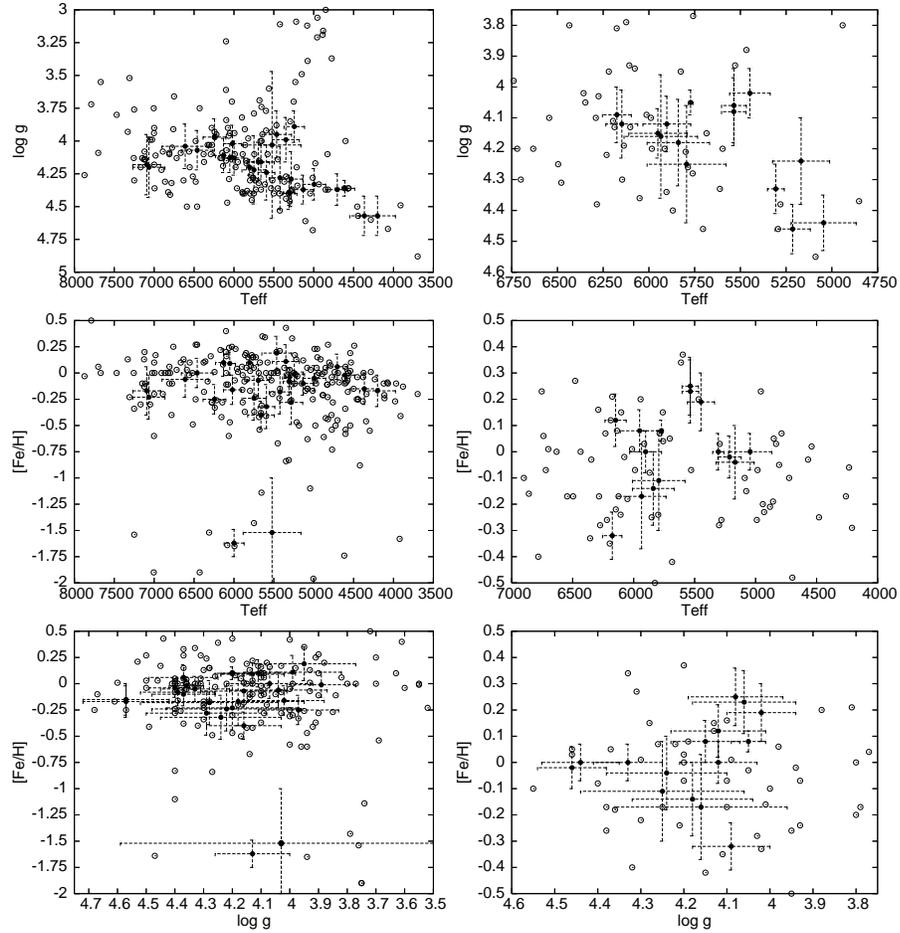}
\caption{
{\it Left:} Distribution of the ELODIE reference stars from Table 7 
(open circles) and the program stars from Table 9 (filled circles) in the 
$T_{\rm eff}$--$\log g$, $T_{\rm eff}$--$\rm [Fe/H]$ and $\log g$--$\rm 
[Fe/H]$ planes. {\it Right:} The same for FRESCO reference stars from 
Table 8 (open circles) and program stars from Table 10 (filled circles). For 
clarity, the diagrams are zoomed to the regions occupied by the program 
stars.
}
\end{center}
\end{figure}

In Fig.\ 5 (left panels), we show the ELODIE reference stars and the program stars, 
plotted in three two-parameter planes, viz., $T_{\rm{eff}}$ -- $\log g$, $T_{\rm{eff}}$ -- 
$\rm [Fe/H]$, and $\log g$ -- $\rm [Fe/H]$. For clarity, the diagrams are zoomed to regions 
occupied by the program stars, so that the reference stars falling outside the panels are 
not shown. 

The right panels of Fig.\ 5 show the FRESCO reference stars (see Table 8) 
and program stars from Table 10. In this figure, the ranges of the 
parameters are those that we have used for computing the astrophysical 
parameters of program stars. 

\begin{figure}
\begin{center}
\epsfbox{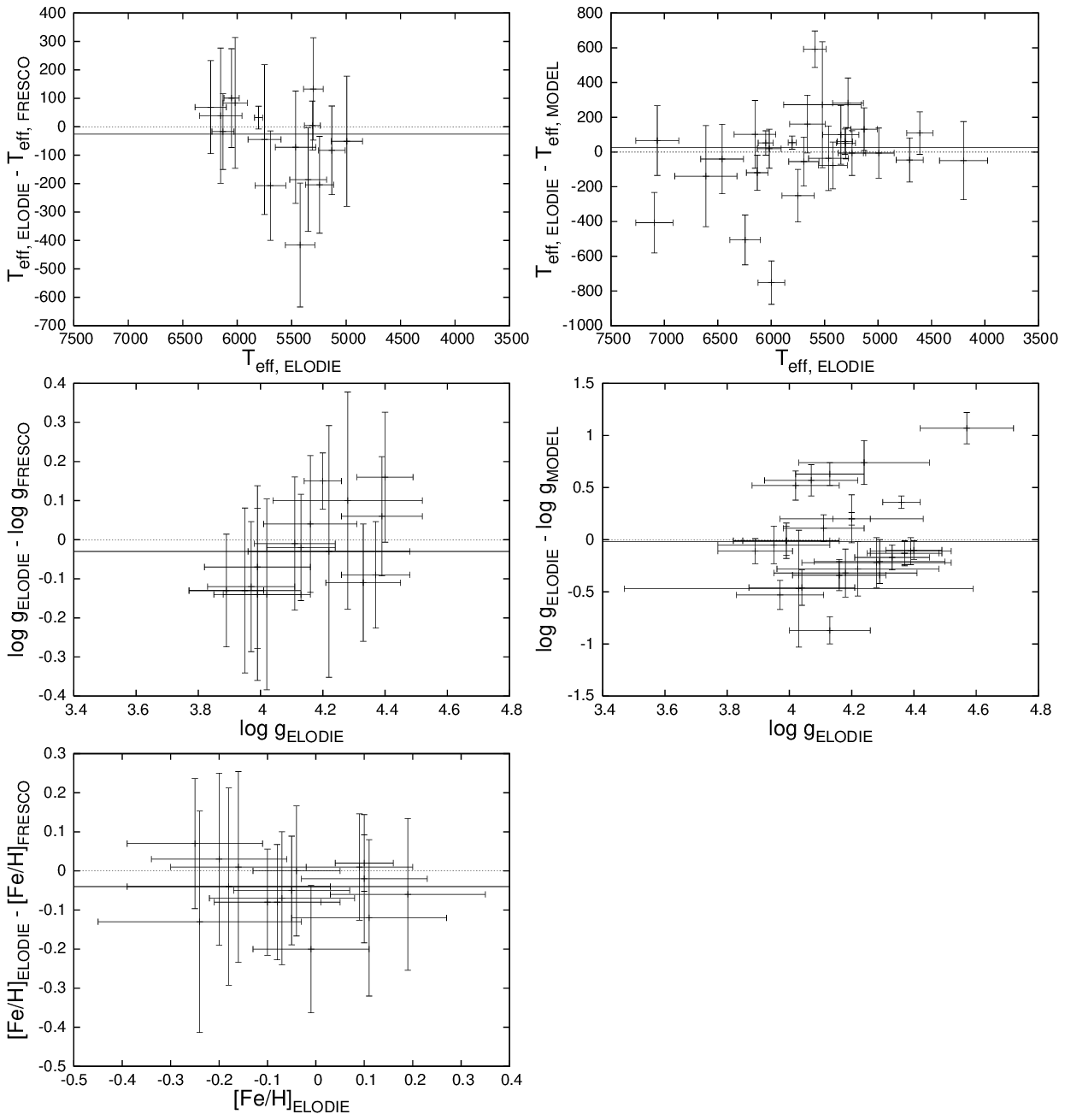}
\caption{Differences, $T_{\rm eff}\ ({\rm ELODIE}) - T_{\rm eff}\
({\rm FRESCO})$ (upper left panel),  $\log g\ ({\rm ELODIE}) - \log g\ 
({\rm FRESCO})$ (middle left panel), and $\rm [Fe/H]$ (bottom left panel), 
between ELODIE and FRESCO determinations, and respective
differences  between ELODIE and model determinations plotted for $T_{\rm 
eff}$ (upper right panel) and $\log g$ (lower right panel).
The dashed lines are zero lines; the solid lines show the mean difference 
between the computed values.}
\end{center} 
\end{figure}

All 16 stars from Table 10 appear also in Table 9. For these stars, the difference 
between the ELODIE and FRESCO based values of the effective temperature, $T_{\rm eff}\ 
({\rm ELODIE}) - T_{\rm eff}\ ({\rm FRESCO})$, is plotted as a function of $T_{\rm eff}\ 
({\rm ELODIE})$ in the upper left panel of Fig.\ 6. The values agree well to within the 
error bars. The largest discrepancy, amounting to about 400\,K, occurs for HIP\,95733. 
If we reject this star, we find the mean difference between $T_{\rm eff}$ determined 
from these two grids to be equal to -27$\pm$28\,K. The cause of the discrepancy in case 
of HIP\,95733 is unclear.

For $\log g$, the difference $\log g\ ({\rm ELODIE}) - \log g\ ({\rm FRESCO})$ is shown 
in the left middle panel.  As can be seen from the figure, the $\log g$ values show 
adequate agreement, although the values based the FRESCO grid are systematically -- 
albeit slightly -- larger than those based on the ELODIE grid. The mean difference 
between $\log g$ determined from these two grids is equal to -0.03$\pm$0.03\,dex.

The ELODIE and FRESCO values of $\rm [Fe/H]$ are compared in the 
bottom panel of Fig.\ 6.  All points agree well; the mean difference 
between $\rm [Fe/H]$ determined from these two grids is equal to
-0.04$\pm$0.07\,dex.

\begin{table}
\begin{center}
\centerline{T a b l e \quad 9}

{Astrophysical parameters and MK spectral types of PATS and the remaining 
stars determined with the use of the ELODIE library. In the last column, 
we list spectral types from the literature}

\vspace{0.3cm}
{\small
\begin{tabular}
{lrrrrrrll} \\
\hline\noalign{\smallskip}
HIP   & $T_{\rm eff}$ & s.d. & $\log g$ & s.d. & $\rm [Fe/H]$ &s.d. 
& MK &lit.\\
\noalign{\smallskip}\hline\noalign{\smallskip}
\multicolumn{9}{c}{PATS}\\
\noalign{\smallskip}\hline\noalign{\smallskip}
93011 & 6460 & 200 & 4.07 & 0.15 &  0.00 & 0.14 &F5IV    &F2\\
94335 & 6152 & 195 & 3.99 & 0.14 & -0.20 & 0.14 &F8V     &F8V+G8V\\
94497 & 4994 & 144 & 4.33 & 0.12 & -0.05 & 0.12 &K2V     &K2\\
94565 & 6052 &  70 & 4.13 & 0.11 &  0.09 & 0.11 &G0IV    &G0\\
94734 & 5804 &  38 & 4.20 & 0.06 &  0.10 & 0.06 &G2V     &---\\
95098 & 6019 & 112 & 4.02 & 0.14 & -0.16 & 0.14 &F8III-IV&F8\\
95733 & 5423 & 135 & 4.28 & 0.24 & -0.18 & 0.21 &G8V     &---\\
95637 & 6611 & 291 & 4.04 & 0.17 & -0.06 & 0.17 &F2III   &F0\\
96634 & 5310 &  73 & 4.39 & 0.13 & -0.08 & 0.13 &K0V     &K0\\
96735 & 5131 & 123 & 4.37 & 0.11 & -0.10 & 0.11 &K2V     &K0\\
97219 & 5301 &  90 & 4.40 & 0.09 & -0.04 & 0.09 &K0V     &G5\\
97337 & 4365 & 184 & 4.57 & 0.15 & -0.15 & 0.15 &K7V     &M0 \\
97657 & 4704 & 127 & 4.37 & 0.12 &  0.06 & 0.12 &K3V     &K2\\
97974 & 5749 & 151 & 4.22 & 0.26 & -0.24 & 0.21 &G0V     &G0\\
98655 & 5282 & 144 & 4.29 & 0.21 & -0.28 & 0.21 &G9V     &K0\\
\noalign{\smallskip}\hline\noalign{\smallskip} 
\multicolumn{9}{c}{The remaining stars}\\
\noalign{\smallskip}\hline\noalign{\smallskip} 
91128 & 4200 & 225 & 4.57 & 0.15 & -0.17 & 0.15 &K7V     &M0\\
92922 & 5464 & 186 & 3.95 & 0.18 &  0.19 & 0.16 &G8IV    &G5\\
94145 & 7093 & 174 & 4.18 & 0.23 & -0.17 & 0.23 &F0III   &F0III\\
94704 & 5522 & 362 & 4.03 & 0.56 & -1.52 & 0.52 &sdG0    &K2 \\
94743 & 7066 & 201 & 4.20 & 0.23 & -0.23 & 0.21 &F2V     &F5\\
94898 & 5244 & 129 & 3.89 & 0.12 & -0.01 & 0.12 &G8IV    &G5\\
95631 & 5349 & 169 & 3.99 & 0.17 &  0.11 & 0.16 &G9IV    &G5\\
95638 & 5695 & 141 & 4.16 & 0.15 & -0.07 & 0.15 &G5V     &G5\\
95843 & 6244 & 143 & 3.97 & 0.14 & -0.25 & 0.14 &F2V     &F2\\
96146 & 6131 & 101 & 4.11 & 0.13 &  0.10 & 0.13 &F8IV-V  &F0\\
97168 & 5661 & 166 & 4.16 & 0.13 & -0.40 & 0.13 &G0V     &---\\
98381 & 4609 & 122 & 4.36 & 0.06 & -0.02 & 0.06 &K4V     &K2\\
98829 & 5592 & 105 & 4.24 & 0.21 & -0.32 & 0.21 &G8V     &G5\\
99267 & 5998 & 125 & 4.13 & 0.13 & -1.62 & 0.13 &sdF2    &F3\\
\noalign{\smallskip}\hline    
\end{tabular} 
}
\end{center}
\end{table}

We conclude that both grids, ELODIE and FRESCO, can 
be safely used for our purpose. The ELODIE grid, however, is more useful 
at this stage because it is sufficiently wide and dense, allowing a better 
determination of astrophysical parameters for stars falling in the coarse 
parts of the FRESCO grid. 

\begin{table}
\begin{center}
\centerline{T a b l e \quad 10}

{Astrophysical parameters of 10 PATS and six remaining 
stars determined with the use of the FRESCO library}

\vspace{0.3cm}
{\small
\begin{tabular}
{lrrrrrr} \\
\hline\noalign{\smallskip}
HIP   & $T_{\rm eff}$ & s.d. & $\log g$ & s.d. & $\rm [Fe/H]$ & s.d.\\
\noalign{\smallskip}\hline\noalign{\smallskip}
\multicolumn{7}{c}{PATS}\\
\noalign{\smallskip}\hline\noalign{\smallskip}
94335 & 6113 & 136 & 4.13 & 0.17 & -0.23 & 0.17 \\
94497 & 5045 & 178 & 4.44 & 0.09 &  0.00 & 0.07 \\
94565 & 5951 & 159 & 4.15 & 0.08 &  0.08 & 0.08 \\
94734 & 5772 &  13 & 4.05 & 0.04 &  0.08 & 0.04 \\
95098 & 5935 & 201 & 4.16 & 0.20 & -0.17 & 0.20 \\
95733 & 5839 & 171 & 4.18 & 0.14 & -0.14 & 0.14 \\
96634 & 5306 &  46 & 4.33 & 0.08 &  0.00 & 0.07 \\
96735 & 5214 &  96 & 4.46 & 0.08 & -0.02 & 0.08 \\
97219 & 5168 & 156 & 4.24 & 0.14 & -0.04 & 0.14 \\
97974 & 5794 & 216 & 4.25 & 0.19 & -0.11 & 0.19 \\
\noalign{\smallskip}\hline\noalign{\smallskip} 
\multicolumn{7}{c}{The remaining stars}\\
\noalign{\smallskip}\hline\noalign{\smallskip} 
92922 & 5536 &  67 & 4.08 & 0.11 &  0.25 & 0.11 \\
94898 & 5448 & 111 & 4.02 & 0.08 &  0.19 & 0.11 \\
95631 & 5535 &  68 & 4.06 & 0.12 &  0.23 & 0.12 \\
95638 & 5902 & 131 & 4.12 & 0.09 &  0.00 & 0.08 \\
95843 & 6175 &  79 & 4.09 & 0.09 & -0.32 & 0.09 \\
96146 & 6148 &  87 & 4.12 & 0.11 &  0.12 & 0.10 \\
\noalign{\smallskip}\hline    
\end{tabular}
}
\end{center}
\end{table}

\vspace{0.5cm}
{\em 4.2. From Model Atmospheres}

\vspace{0.5cm}
In addition, we derived global
atmospheric parameters with the use of
model atmospheres. These computations
were run for spectrograms measured at the ORO and FLWO. Here, we used 
one-dimensional correlations to identify the template in the library of 
synthetic spectra that gives the best match with the observed spectrum. 
We chose the template that gave the highest peak correlation value 
averaged over all the observed spectra for each program star. We assumed 
solar metalicity for all but five high proper-motion stars, namely, 
HIP\,98655, HIP\,97168, and HIP\,98829, for which we used $\rm [Fe/H]$ = -0.5, and 
HIP\,94704 and HIP\,99267, for which we used $\rm [Fe/H]$ = -1.5, and then we 
solved for effective temperature and surface gravity. 
The composite spectrum of HIP\,94335
was analyzed with a two-dimensional correlation 
technique TODCOR (Zucker \& Mazeh~1994, and Torres et al.~2002). Grids of 
solutions were run using the CfA library of synthetic spectra as templates 
in order to find the individual templates for the primary and secondary 
that gave the best match to the observed spectra. Solar metalicity was 
assumed for this analysis, and surface gravities of $\log g = 4.0$ and 4.5 
were adopted for the primary and secondary, respectively.  The best match 
to the observed spectra was achieved using a template with $T_{\rm eff} = 
6060 \pm 50$ and $v \sin i = 30 \pm 2$ for the primary and $T_{\rm eff} = 
5390 \pm 250$ and $v \sin i = 24 \pm 4$ for the secondary. 
In Table 11, we list the parameters obtained for 14 PATS and 14 remaining 
stars. 

The difference between effective temperature obtained with the use of
ELODIE grid and the model atmospheres is plotted as a function of 
$T_{\rm eff}\ 
({\rm ELODIE})$ in the upper right panel of Fig.\ 6. In most cases, the 
values agree to within the error bars. Only for HIP\,94145, HIP\,95843, HIP\,98829,
and HIP\,99267 there are high discrepancies.

For HIP\,99267, $T_{\rm eff}$ computed from model atmospheres is 752\,K higher 
than that obtained from ELODIE grid, which is equal to 5998\,K. It is also
higher than other determinations
that can be found in the literature, and which range from 5650\,K (Carney 
et al.~1997) to 5857\,K (Laird et al.~1988).
For HIP\,94145, classified in this paper and by Macrae~(1952)  to
F0\,III, the effective temperature obtained from model atmospheres is 
around 400\,K higher than $T_{\rm eff}$ derived form ELODIE grid, and is not 
consistent with $T_{\rm eff}$ expected for a typical F0 giant (see, e.g., 
Alonso et al.~1999).
For HIP\,98829, classified in this paper to G8V and by Dieckvoss \& 
Heckmann~(1975) to G5, the effective temperature obtained from model 
atmospheres is around 600\,K lower than the one derived from ELODIE 
grid, and is typical for a K2V star.
For HIP\,95843, that is a slightly metal--deficient star classified to
F2V in this paper and to F2 by Dieckvoss \& Heckmann~(1975), the $T_{\rm eff}$ 
derived from model atmospheres is around 500\,K higher than 
the one derived from ELODIE grid.  
The source of these discrepancies is not clear. After rejecting these
four stars, the mean difference between $T_{\rm eff}$ determined from 
ELODIE grid and the model atmospheres is equal to +26$\pm$122\,K.
 
For $\log g$, the difference $\log g\ ({\rm ELODIE}) - \log g\ ({\rm 
MODEL})$ is shown in the right lower panel of Fig.\ 6. These values
show good overall agreement with a mean equal to -0.02$\pm$0.43\,dex.
The highest discrepancies occur for HIP\,91128 and HIP\,99267.

\begin{table}
\begin{center}
\centerline{T a b l e \quad 11}

{Astrophysical parameters of 14 PATS and 14 remaining
stars determined with the use of model atmospheres\\ (as 
elsewhere in this paper HIP\,94335 is counted as one 
star)
}
\vspace{0.3cm}
{\small
\begin{tabular}
{lrrr|lrrr} \\
\hline\noalign{\smallskip}
HIP   &$T_{\rm eff}$ & $\log g$ & $\rm [Fe/H]$ &
HIP   &$T_{\rm eff}$ & $\log g$ & $\rm [Fe/H]$\\
\noalign{\smallskip}\hline\noalign{\smallskip}
\multicolumn{4}{c}{PATS}&\multicolumn{4}{c}{The remaining stars}\\
\noalign{\smallskip}\hline\noalign{\smallskip}
 93011   & 6500 & 3.5 & 0.0 & 91128 & 4250 & 3.5 & 0.0\\
 94335-1 & 6060 & 4.0 & 0.0 & 92922 & 5500 & 4.0 & 0.0\\
 94335-2 & 5390 & 4.5 & 0.0 & 94145 & 7500 & 4.5 & 0.0\\
 94497   & 5000 & 4.5 & 0.0 & 94704 & 5250 & 4.5 &-1.5\\
 94565   & 6000 & 3.5 & 0.0 & 94743 & 7000 & 4.0 & 0.0\\
 94734   & 5750 & 4.0 & 0.0 & 94898 & 5250 & 4.0 & 0.0\\
 95098   & 6000 & 3.5 & 0.0 & 95631 & 5250 & 4.0 & 0.0\\
 95637   & 6750 & 4.5 & 0.0 & 95638 & 5750 & 4.5 & 0.0\\
 95733   & 5500 & 4.5 & 0.0 & 95843 & 6750 & 4.5 & 0.0\\
 96634   & 5250 & 4.5 & 0.0 & 96146 & 6250 & 4.0 & 0.0\\
 96735   & 5000 & 4.5 & 0.0 & 97168 & 5500 & 4.5 &-0.5\\
 97219   & 5250 & 4.5 & 0.0 & 98381 & 4500 & 4.0 & 0.0\\
 97657   & 4750 & 4.5 & 0.0 & 98829 & 5000 & 3.5 &-0.5\\
 97974   & 6000 & 4.5 & 0.0 & 99267 & 6750 & 5.0 &-1.5\\
 98655   & 5000 & 4.5 &-0.5 & \\
\noalign{\smallskip}\hline
\end{tabular}
}
\end{center}
\end{table}

\vspace{0.5cm} 
{\bf 5. Projected Rotational Velocity}

\vspace{0.5cm} 
In Table 12, we list projected rotational velocities of the program stars.  For 
each star, two values are given, one determined from a grid of Kurucz 
model spectra (columns headed ``model''), and another, determined with 
the Full Width at Half Maximum (FWHM) method. 

We have used the Kurucz model spectra (see Sect.\ 2) for stars 
observed with the MMT, Wyeth, and Tillinghast telescopes. In this method, 
each observed spectrum is compared with a library of synthetic spectra 
using correlation techniques described already in Sec.\ 4.2. A typical
standard deviation of these determinations is equal to 1 or 2 km\,s$^{-1}$.

For stars observed with FRESCO, we have determined $v\sin i$ using the FWHM method for 
each order of the echelle spectrum. We excluded orders containing very broad spectral 
lines that can affect the shape of the cross-correlation function. As templates, we used 
a grid of rotationally broadened spectra of a non-rotating star having a similar $T_{\rm 
eff}$, $\log g$, and $\rm [Fe/H]$ as the program star. An upper limit of 5~km\,s$^{-1}$ 
has been estimated according to the instrumental resolution of the spectrograms. The 
values of $v\sin i$ larger than this limit are listed in Table 12 together with their 
standard deviations. 

As can be seen from Table 12, the  values of $v\sin i$ obtained from the two separate 
sets of data by means of the above-mentioned two methods agree well. The only exception 
is HIP\,99267, an extremely metal-poor star with so few lines that it was difficult to 
obtain a precise value of $v\sin i$ using the FWHM method.

\begin{table}
\begin{center}
\centerline{T a b l e \quad 12}

{Projected rotational velocities determined from a grid of Kurucz model spectra 
and from the FWHM method}

\vspace{0.3cm}
{\small
\begin{tabular}
{lrrr|lrrr} \\
\hline\noalign{\smallskip}
HIP   &$v\sin i$&$v\sin i$&s.d.&HIP   &$v\sin i$&$v\sin i$&s.d.\\
  &[km\,s$^{-1}$]  &[km\,s$^{-1}$] &[km\,s$^{-1}$]    &   &[km\,s$^{-1}$]   &[km\,s$^{-1}$] &[km\,s$^{-1}$]   \\
  &\tiny (model) &\tiny (FWHM) &\tiny (FWHM)    &  &\tiny (model)  &\tiny (FWHM) & \tiny (FWHM)  \\
\noalign{\smallskip}\hline\noalign{\smallskip}
\multicolumn{4}{c}{PATS}&\multicolumn{4}{c}{The remaining stars}\\
\noalign{\smallskip}\hline\noalign{\smallskip}
93011   & 30.0 & 28.8   & 0.8 & 91128   &  1.0 & $<$5.0 & \\
94335-1 & 29.5 & 32.7   & 1.2 & 92922   &  1.5 & $<$5.0 & \\
94335-2 & 25.0 & 28.0   & 2.0 & 94145   & 12.3 & 14.3   & 1.1 \\
94497   &  0.0 & $<$5.0 &     & 94704   &  1.5 & $<$5.0 & \\
94565   &  7.3 & 7.4    & 1.0 & 94743   & 15.7 & 14.7   & 0.6 \\
94734   &  1.5 & $<$5.0 &     & 94898   &  3.0 & $<$5.0 & \\    
95098   &  3.0 & $<$5.0 &     & 95631   &  2.5 & $<$5.0 & \\
95637   & 28.5 & 29.5   & 0.9 & 95638   &  1.5 & $<$5.0 & \\
95733   &  0.5 & $<$5.0 & & 95843   & 12.6 & 17.1   & 1.0 \\      
96634   &  1.5 & $<$5.0 & & 96146   & 33.5 & 35.6   & 1.6 \\      
96735   &  1.5 & $<$5.0 & & 97168   &  1.5 & $<$5.0 & \\          
97219   &  0.0 & $<$5.0 & & 98381   &  1.5 & $<$5.0 & \\          
97337   &  --- & $<$5.0 & & 98829   &  3.0 & $<$5.0 & \\          
97657   &  0.0 & $<$5.0 & & 99267   &  7.6 & 10.4   & 2.0\\       
97974   &  1.5 & $<$5.0 & &\\                                     
98655   &  1.5 & $<$5.0 & &\\                                     
\noalign{\smallskip}\hline                                    
\end{tabular}
}
\end{center}
\end{table}

\vspace{0.5cm}
\centerline{\bf 6. Summary}

\vspace{0.5cm}
We present spectroscopic observations of 15 best candidates for Kepler 
primary asteroseismic targets (PATS) and of 14 other program stars. 
The observations were carried out at the {\it M.G. 
Fracastoro\/} station of the Catania Astrophysical Observatory, the 
Oak Ridge Observatory, Harvard, Massachusetts, and the F.L.\ Whipple 
Observatory, Mount Hopkins, Arizona.  

We find that all PATS have solar-like metalicity or are slightly 
metal-deficient. They range from 
early F-type to late K-type, so that we expect all of them to show
solar-like oscillations. Four PATS we classify as subgiants or giants. 
These stars are particularly interesting from the asteroseismic point 
of view because the predicted amplitude of solar-like oscillations in the
evolved stars is expected to be higher than in dwarfs (see, e.g., 
Kjeldsen \& Bedding~1995).

From all spectrograms, we derive the radial velocities. The results are given in Table 
2, available in electronic form from the Acta Astronomica Archive (see the cover page). 
The spectrograms obtained at the {\it M.G.  Fracastoro\/} station of the Catania 
Astrophysical Observatory are used to determine the effective temperature, surface 
gravity, metalicity, and MK type by means of the ROTFIT code (Frasca et al. 2003, 2006). 
The spectrograms obtained at the Oak Ridge Observatory and the F.L.\ Whipple Observatory 
are used to determine the effective temperature and surface gravity by means of a 
two-dimensional correlation technique TODCOR (Zucker \& Mazeh~1994 and Torres et 
al.~2002).  The results obtained from these two different methods applied to two 
different data sets and given in Tables 9, 10, and 11, agree well in most cases. We also 
estimate the projected rotational velocity from two separate sets of data using two 
independent methods (see Table 12), obtaining good agreement.

We discuss in some detail three program stars that show variable 
radial-velocity, viz., HIP\,94335, HIP\,94734, and HIP\,94743. For 
HIP\,94335 = FL Lyr, one of the PATS and a well-known Algol-type 
eclipsing variable and a double-lined spectroscopic binary,
we find close agreement between orbital 
elements computed from our data (see Table 4 and Fig.\ 1) and those of 
Popper et al.\ (1986). For this star precise
determination of mass and radius exist in the literature (see Torres
et al.~2006 and references therein) that can be used 
either as an additional constraint for pulsational modeling or as a 
test of its results. 

HIP\,94734, which is one of PATS, and HIP\,94743 = V\,2077 Cyg, which 
does not fall onto active pixels of Kepler CCDs, we have discovered to 
be single-lined systems. For both, we derive orbital elements 
(see Tables 5 and 6, and Figs.\ 2 and 3).  In addition, using our value of 
the orbital period and Hipparcos epoch photometry (ESA 1977), we 
demonstrate that HIP\,94743 = V\,2077 Cyg is a detached eclipsing binary 
(see Fig.\ 4). 

Our data do not confirm the spectroscopic-binary status of HIP\,97219 = 
HD 187055, reported to be variable in radial velocity by Nordstr\"{o}m et 
al.~(2004), and HIP\,99267 = G125-64, suspected to be a spectroscopic 
binary by Carney and Latham (1987). 

\vspace{0.5cm}{\bf Acknowledgments.} 
This work was supported by MNiSW grant N203 014 31/2650, the University of 
Wroc{\l}aw grants 2646/W/IA/06 and 2793/W/IA/07, the Italian government 
fellowship BWM-III-87-W{\l}ochy/ED-W/06 and the Socrates-Erasmus 
Program ``Akcja 2'' 2006-2007, contract No.\ 33. J.M.-\.Z.\ thanks the 
Danish Natural Science Research Council, the Italian National Institute 
for Astrophysics (INAF), the University of Catania, and the University of 
Wroc{\l}aw for the financial support.

We acknowledge the partial support from the Kepler mission under 
cooperation agreement NCC2-1390 (D.W.L., PI).

\vspace{0.5cm}
\centerline{REFERENCES}
\vspace{0.3cm}
{\small

{Alonso, A., Arribas, S., and Martnez-Roger, C.} {1999} {Astron.\ Astrophys.\ Suppl.\ Ser.} {140} {261}

{Borucki, W.J., Koch, D.G., Dunham, E.W., and Jenkins, J.M.} {1997} {ASP Conf.\ Ser.} {vol. 119} {p.153} 
{ed.\ David Soderblom}

{Carney, B.W., and Latham, D.W.} {1987} {Astron.\ J.} {93} {116}

{Carney, B.W., Wright, J.S., Sneden, C., Laird, J.B., Aguilar, L.A., and Latham, D.W.} 
{1997} {Astron.\ J.} {114} {363}

{Christensen-Dalsgaard, J.} {2004} {Solar Physics} {220} {137}

{Dieckvoss, W., and Heckmann, O.} {1975} {AGK3 Catalogue}

{ESA} {1997} {``The Hipparcos and Tycho Catalogues''} {ESA SP-1200}

{Frasca, A., Alcal\`a, J. M., Covino, E., Catalano, S., Marilli, E., and Paladino, R.} 
{2003} {Astron.\ Astrophys.} {405} {149} 

{Frasca, A., Guillout, P., Marilli, E., Freire Ferrero, R., Biazzo, K., and Klutsch, A.} 
{2006} {Astron.\ Astrophys.} {454} {301}

{Johnson, H.L.} {1966} {Ann.\ Rev.\ Astron.\ Astrophys.} {4} {193}

{Katz, D., Soubiran, C., Cayrel, R., Adda, M., and Cautain, R.} {1998} 
{Astron.\ Astrophys.} {338} {151}

{Kazarovets, A.V., Samus, N.N., Durlevich, O.V., Frolov, M.S., Antipin, S.V., Kireeva, 
N.N., and Pastukhova, E.N.} {1999} {IBVS} {No.\ 4659}

{Kjeldsen, H., and Bedding T.} {1995} {Astron.\ Astrophys.} {293} {87}

{Laird, J.B., Carney, B.W., and Latham, D.W.} {1988} {Astron.\ J.} {95} {1843}

{Latham, D.W., Mazeh, T., Stefanik, R.P., Davis, R.J, Carney, B.W., Krymolowski, Y., 
Laird, J.B., Torres, G., and Morse, J.A.} {1992} {Astron.\ J.} {104} {774}

{Latham, D.W., Stefanik, R.P., Torres, G., Davis, R.J., Mazeh, T., Carney, B.W., 
Laird, J.B., and Morse, J.A.} {2002} {{Astron.\ J.} {124} {1144}

{Macrae, D.A.} {1952} {Astrophysical J.} {116} {592}
 
{Molenda-\.Zakowicz, J., Arentoft, T., Kjeldsen, H., and Bonanno, A.} {2006} 
{Proc.\ SOHO 18/GONG 2006/HELAS I}, {CDROM, p.\ 110.1}

{Nordstr\"{o}m, B., Mayor, M., Andersen, J., Holmberg, J., Pont, F., J{\o}rgensen, 
B.R., Olsen, E.H., Udry, S., and Mowlavi, N.} {2004} {Astron.\ Astrophys.} {418} {989} 

{Popper, D.M., Lacy, C.H., Frueh, M.L., and Turner, A.E.} {1986} {Astron.\ J.} {91} {383}

{Prugniel, Ph., and Soubiran, C.} {2001} {Astron.\ Astrophys.} {369} {1048}

{Soubiran, C., Katz, D., and Cayrel, R.} {1998} {Astron.\ Astrophys.\ Suppl.} {133} {221}

{Torres, G., Boden, A.F., Latham, D.W., Pan, W., and Stefanik, R.} {2002} 
{Astron.\ J.} {124} {1716}

{Torres, G., Lacy, C.H.,Marschall, L.A., Sheets, H.A., and Mader, J.A.} {2006} 
{Astrophys.\ J.} {640} {1018}

{Udry, S., Mayor, M., Maurice, E., Andersen, J., Imbert, M., Lindgren, H., Mermilliod, 
J.-C., Nordstr\"{o}m, B., and Pr\'evot, L.} {1999} {IAU Coll.} {170} 
{ASP Conf.\ Ser.} {185} {p.\ 383}

{Zucker, S., and Mazeh, T.} {1994} {Astrophys.\ J.} {420} {806}
}

\end{document}